\documentclass[aps,pra,reprint,amsmath,amssymb,twocolumn,preprintnumbers,superscriptaddress,10pt,longbibliography,bibnotes]{revtex4-2}
\usepackage[utf8]{inputenc}                                             % Input encoding

\usepackage{xcolor}                                                     % Colors
\usepackage{amsmath,amssymb}                                            % Math
\usepackage{amsthm}                                                     % Theorem environments
\usepackage[italicdiff]{physics}                                        % Physics
\usepackage{dsfont}                                                     % Double stroke math symbols
\usepackage{graphicx}                                                   % Include figure files
\usepackage{tikz}                                                       % Tikz figures
\usepackage{pgfplots}                                                   % PGF plots
\pgfplotsset{compat=1.14}
\usetikzlibrary {arrows.meta}

\usepackage{hyperref}                                                   % Hyperlinks

\usepackage{xr}
\makeatletter
\newcommand{\raisemath}[1]{\mathpalette{\raisem@th{#1}}}
\newcommand{\raisem@th}[3]{\raisebox{#1}{$#2#3$}}
\makeatother

\newcommand*{\lp}{\mathopen{}\left}                                     % Left parenthesis
\newcommand*{\rp}{\right}                                               % Right parenthesis
\newcommand*{\hc}{\text{H.c.}}                                          % Hermitian conjugate
\newcommand{\vekh}[1]{\hat{{\boldsymbol{\mathbf{#1}}}}}

\renewcommand{\bra}[1]{\big\langle #1 \big|}
\renewcommand{\ket}[1]{\big| #1 \big\rangle}

\newcommand*{\smallket}[1]{| #1 \rangle}                        % Ket
\newcommand*{\smallbra}[1]{\langle #1 |}                        % Bra

\newcommand*{\supplementary}{Supplemental Material}                     % Supplementary section reference
\newcommand*{\regi}{\text{I}} % Label region to the left
\newcommand*{\regii}{\text{II}} % Label region right

\begin{document}

%===== Preprint ===============
%\preprint{APS/123-QED}

%===== Title ==================
%% Force line breaks with \\
\title{A truncated photon}

%===== Authors ================
\author{Isak Cecil Onsager Rukan}
\affiliation{Department of Physics, University of Oslo, NO-0316 Oslo, Norway}

\author{Jan Gulla}
\affiliation{Department of Physics, University of Oslo, NO-0316 Oslo, Norway}
%\affiliation{Department of Technology Systems, University of Oslo, NO-0316 Oslo, Norway}

\author{Johannes Skaar}
\email{johannes.skaar@fys.uio.no}
\affiliation{Department of Physics, University of Oslo, NO-0316 Oslo, Norway}

%===== Date ===================
\date{\today}

%===== Abstract ===============
\begin{abstract}
An elementary particle such as a photon cannot be cut in two pieces. Still it must be possible to truncate a photon with an optical shutter. The result is neither another photon nor a mix of a photon and a vacuum. Instead it is a superposition and mix of photon numbers up to infinity. This state is rather complicated, but nevertheless locally equivalent to a single photon or vacuum to the left and right, respectively, of a narrow transition region. \end{abstract}

%===== Keywords ===============
%% Use showkeys class option if keyword display desired
%\keywords{single photons, causality, photon localization, single photon generation}

%===== Maketitle ==============
\maketitle

A photon, being an elementary particle, cannot be cut in two halves. Still, it is clearly possible to remove a part of an optical pulse, e.g. using an optical shutter. If a photon is truncated in this way, what is the resulting state? Despite being a simple question, it appears that it has not been asked before. In addition to being an interesting question from a fundamental point of view, we will argue below that it also has implications for our understanding of localized states and local equivalence in quantum field theory.

The simplest case of a forward-propagating photon $\smallket{1_\xi}$ in a mode $\xi(\omega)$ incident on a time-independent mirror with transmissivity $p>0$ is well-known. Instead of being ``cut'', the photon is simply transformed into a superposition of forward- and backward-propagating modes $1$ and~$2$:
\begin{equation}\label{1_xi_beamsplitter}
    \ket{1_\xi} \mapsto \sqrt{p} \ket{1_1} \ket{0_2} + \sqrt{1 - p} \ket{0_1} \ket{1_2}.
\end{equation}
It is then perhaps tempting to think the same model can be used to add time dependence to the mirror. Let $\xi(t)$ be the inverse Fourier transform of the mode $\xi(\omega)$. With a perfectly reflecting mirror initially present, a part $\xi_2(t)$ of the incoming photon is reflected to the backward mode, and suddenly removing the mirror lets the rest of the pulse $\xi_1(t) = \xi(t) - \xi_2(t)$ through, so that
\begin{equation}\label{1_xi_time}
    \ket{1_\xi} \mapsto \sqrt{p} \ket{1_{\xi_1}} \ket{0_{\xi_2}} + \sqrt{1 - p} \ket{0_{\xi_1}} \ket{1_{\xi_2}}.
\end{equation}
Here $p$ is a constant resulting from normalizing $\xi_1(t)$ and $\xi_2(t)$. Tracing out the backward mode, we get the forward-propagating mixed state
\begin{equation}\label{1_xi_time_forward}
    \ket{1_\xi} \mapsto  p \ket{1_{\xi_1}}\bra{1_{\xi_1}} + (1-p) \ket{0_{\xi_1}}\bra{0_{\xi_1}}.
\end{equation} 

In particular, this seems to suggest that we can use the mirror to truncate the photon into a mode with some compact support $\xi_1(t)$. This, however, cannot be correct. Single photons have infinite tails \cite{knight1961,bialynicki-birula1998}, so the right-hand side of \eqref{1_xi_time_forward} cannot represent a state that looks like vacuum outside the support of $\xi_1(t)$. Indeed, the familiar methods that work for modes in frequency space cannot simply be employed for temporal modes $\xi_1(t)$, although they can be approximately valid for narrow-bandwidth pulses~\cite{loudon2000}. 

So then, what does a truncated photon state look like? Surprisingly, we will find that a truncated photon is a complicated state involving photon numbers up to infinity. In other words, by ``cutting away part of the photon'' with a shutter, we effectively create a bunch of new photons. In principle, this is the case even if the shutter is turned off slowly.

The creation of photons can be understood as a consequence of the breaking of time-translation invariance; accordingly, by Noether’s theorem, the field energy is not conserved. In fact, from the extensive literature on the dynamical Casimir effect \cite{moore1970,dewitt1975,fulling1976,yablonovitch1989, dodonov1993,law1994,cirone1999,braunstein2005} with moving or changing mirrors or dielectrics, it is well known that the transformed vacuum state contains photons. The dynamical Casimir effect is even observed experimentally~\cite{wilson_observation_2011}. Here, of special relevance are the works by Cirone and Rzaszewski~\cite{cirone1999}, and Braunstein \cite{braunstein2005} in which a mirror in a cavity is removed. These studies were performed in the Heisenberg picture to calculate the expected number of photons in the initial vacuum state.

We want to determine how an incident single photon is transformed by truncation, using a time-dependent shutter. This turns out to be somewhat tricky in two ways: First, we will need the detailed form of the state, expressed with the transformed ladder operators. Since the system is inhomogeneous in space, different wavevectors are coupled together. Second, an instantaneous removal of the shutter, which we treat for simplicity in what follows, produces an infinite number of photons. Thus, we must support our treatment with an analysis of gradual removal, which involves a non-local transformation in time.

In the following, we define the setup, deduce the Bogoliubov transformation that governs the shutter, and describe the form of the resulting state. This includes an estimate of the fidelity to an exact, forward-propagating photon state, expressed from the size of the removed tail. 
Finally, we discuss how the truncated photon illuminates concepts of local equivalence and localization in quantum field theory.

\begin{figure}[t]
    \begin{center}
    \begin{tikzpicture}[scale=1.0]
        % Horizontal axis
        \draw[->] (-3.5,0) -- (2.5,0) node[below right] {$x$};
        
        % Reflector
        \draw[very thick] (0,-0.3) -- (0,1);
        \draw[very thick] (0,-1.2) -- (0,-1.8);
        %\node[above] at (0,1.5) {Reflector};
        
        % Incoming wave (with exp(-x) photon tails)
        \draw[blue,thick,smooth] plot[domain=-5:0,samples=200,scale=0.7] (\x,{1.8*exp(-(\x+1.6)^2/(1+abs(\x+1.6)^(1.001)))});
        \draw[blue,arrows={-Stealth[inset=0pt,angle=50:6pt]}] (-1.5, 1.6) -- (-0.7, 1.6); 
        
        % Labels, \ket{1} and \ket{0}
        \node[blue] at (-1.1,1.5/2) {\(\ket{1}\)};
        \node[black] at (1.1,1.5/2) {\(\ket{0}\)};
        
        % Reflected wave
        \draw[blue,thick,dashed,smooth] plot[domain=-5:0,samples=200,scale=0.7](\x,{1.8*exp(-(-\x+1.6)^2/(1+abs(-\x+1.6)^(1.001))});
        \draw[blue,dashed,arrows={-Stealth[inset=0pt,angle=50:6pt]}] (-0.7, -0.4) -- (-1.5, -0.4); 
        
        % Tick marks
        \draw (0,-0.5) node {$0$};

        %\draw (-4.3,0.2) node {$t=0^-$:};
        \draw (-4.1,-1.5) node {$t<0$:};
        \draw (-4.1,-2.5) node {$t>0$:};
        \draw (-1.9,-1.5) node {$E^\regi(t-x) - E^\regi(t+x)$};
        \draw (2,-1.5) node {$E^{\regii}(t+x) - E^{\regii}(t-x)$};
        \draw (0,-2.5) node {$E_a(t-x) + E_b(t+x)$};
    \end{tikzpicture}
    \caption{An incident photon propagating to the right gets reflected in $x=0$. The solid blue line is the expected energy density of the right-going part; the dashed line is the left-going part. At $t=0$ the reflector is removed. The quantum field for $t<0$ is disconnected into the two regions $x<0$ and $x>0$ due to the reflector. The field for $t>0$ is the usual superposition of forward- and backward-propagating parts.} 
    \label{fig:setup} 
    \end{center}
\end{figure}
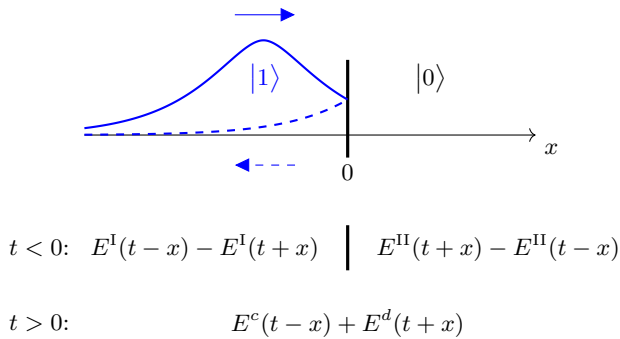

\emph{Setup and model.}
We let the shutter be realized as a time-dependent, perfect reflector in a vacuum. The reflector is effective for negative times $t<0$; at $t=0$, it is removed instantly. We imagine an exact single photon incident from the left (see Fig. \ref{fig:setup}). The photon meets the reflector and a part of the leading tail gets reflected before $t=0$. 

We consider forward- and backward-propagating modes in one dimension, and a single polarization. 
The electromagnetic field is therefore transverse to the propagation axis $x$. We first describe the field for $t<0$ and $t>0$ separately, before connecting by temporal continuity at $t=0$. This leads to Bogoliubov transformations for the ladder operators. Then we define the initial state and express it with the transformed ladder operators, corresponding to the situation after the reflector has been removed.

For $t>0$ the electric field is given by the standard expression
\begin{align}\label{Efieldw}
	E(x,t>0) &= \int_{-\infty}^\infty dk \mathcal E(\omega) a(k)e^{-i\omega t+ikx} + \hc
\end{align}
where $\omega = |k|$ and $\mathcal E(\omega) = K\sqrt\omega$ with a constant $K>0$. For simplicity, we have set the speed of light to unity. Define $b(k)=a(-k)$, $k>0$, which is the annihilation operator for backward-propagating modes. We label the forward- and backward-propagating  parts of \eqref{Efieldw} by
\begin{subequations}\label{EcEddef}
\begin{align}
    E_a(t) &= \int_{0}^{\infty} d\omega \mathcal E(\omega) \left[a(\omega) e^{-i\omega t} + a^{\dag}(\omega)e^{i\omega t}\right], \label{E_a_def} \\
    E_b(t) &= \int_{0}^{\infty} d\omega \mathcal E(\omega) \left[b(\omega) e^{-i\omega t} + b^{\dag}(\omega)e^{i\omega t}\right], \label{E_b_def}
\end{align}
\end{subequations}
respectively, such that the field \eqref{Efieldw} can be written as
\begin{align}
	E(x,t>0) = E_a(t-x) + E_b(t+x).
	\label{Efieldwc}
\end{align}

For $t<0$, when the reflector is effective, the field gets reflected at $x=0$. Since the reflector is assumed perfect, the field is disconnected into independent fields on each side, denoted by superscripts $\regi$ and $\regii$. We let the reflector have $\pi$ phase shift from both the left and the right. Similarly to \eqref{Efieldw}, the electric field can then be written
\begin{align}\label{fieldIII}
	E(x,t<0) = 
	\begin{cases}
		E^\regi(t-x) - E^\regi(t+x), & x<0, \\
		E^{\regii}(t+x) - E^{\regii}(t-x), & x>0,
	\end{cases} 
    % \quad t < 0,
\end{align}
% where $E^{I}$ and $E^{II}$ are of the same structural form as \eqref{E_a_def} and \eqref{E_b_def}, only with the ladder operators $a^{\regi}$ and $b^{\regii}$ instead, respectively.
where
\begin{subequations}
\begin{align}
	E^\regi(t) &= \int_{0}^{\infty} d\omega \mathcal E(\omega) \left[a^{\regi}(\omega) e^{-i\omega t} + a^{\regi^{\dag}}(\omega)e^{i\omega t}\right], \\
	E^{\regii}(t) &= \int_{0}^{\infty} d\omega \mathcal E(\omega) \left[b^{\regii}(\omega) e^{-i\omega t} + b^{\regii^{\dag}}(\omega)e^{i\omega t}\right].
\end{align}
\end{subequations}
Note that in the absence of the mirror, the forward- and backward-propagating parts \eqref{Efieldwc} are independent, with different ladder operators. On the other hand, when the mirror is present, the forward- and backward-propagating parts \eqref{fieldIII} are associated with the same ladder operator, due to reflection, but there is in return one independent mode for each region I and II. We therefore need no $b^\regi(\omega)$ or $a^\regii(\omega)$.

\emph{Bogoliubov transformations.} In time-independent situations, we typically model devices using transformations of the mode operators \cite{leonhardt2003}. With time dependence, we must go to first principles and impose classical boundary conditions on the field, interpreted as operator equations~\cite{moore1970}. Noting that there is a vacuum away from the reflector, the temporal boundary conditions connecting the fields for $t<0$ and $t>0$ are
\begin{subequations}\label{boundt}
\begin{align}
	E(x,0^-) &= E(x, 0^+), \label{Eboundt}\\
	B(x,0^-) &= B(x, 0^+), \label{Bboundt}
\end{align}
\end{subequations}
where $B(x,t)$ is the magnetic field. Eq. \eqref{Eboundt} is a consequence of requiring finite displacement current in Ampère--Maxwell's law. Similarly, \eqref{Bboundt} results from Faraday's law 
% \footnote{For a thorough treatment of the temporal continuity at $x=0$, see the \supplementary \ at [],}.
\footnote{For a thorough treatment of the temporal continuity at $x=0$, see the \supplementary.}.

In vacuum, Faraday's law and Ampère--Maxwell's law give
% \begin{subequations}
\begin{align}
	\frac{\partial B}{\partial t} = - \frac{\partial E}{\partial x}, 
    %\\
    \quad
	\frac{\partial E}{\partial t} = - \frac{\partial B}{\partial x}, \label{faraday}
\end{align}
% \end{subequations}
respectively. Comparing to \eqref{Efieldwc}, the magnetic field 
\begin{equation}
	B(x,t>0) = E_a(t-x) - E_b(t+x),
    %  \quad t>0,
\end{equation}
is consistent with \eqref{faraday}. Also for $t<0$, we find the magnetic field from the electric field by changing the sign of the backward-propagating term. Applying the boundary conditions \eqref{boundt}, we obtain
\begin{subequations} \label{transffieldinv}
	\begin{align}
		E^\regi(x) &= 
			\theta(x)E_a(x) - \theta(-x)E_b(x), \label{trfIInva}\\
		E^{\regii}(x) &= 
			-\theta(-x)E_a(x) + \theta(x)E_b(x),
	\end{align} 
\end{subequations}
where $\theta(x)$ denotes the Heaviside step function.

Let $\xi(\omega)$ be any spectrum such that $\| \xi \| =1$, where ${\|\cdot\|}$ denotes the $L^2$ norm. We define forward and backward creation operators with spectrum $\xi$ by
\begin{equation}\label{axidint}
    a_\xi^\dagger = \int_0^\infty d\omega\xi(\omega)a^\dagger(\omega), \quad b_\xi^\dagger = \int_0^\infty d\omega\xi(\omega)b^\dagger(\omega), 
\end{equation}
respectively. Note that \eqref{axidint} is defined as integrals over positive frequencies only. Taking the Fourier transform of \eqref{transffieldinv}, and defining 
\begin{equation}\label{xix}
    \xi(x) = \frac{1}{2\pi} \int_0^{\infty} d\omega \frac{\xi(\omega)}{\mathcal E(\omega)} e^{i\omega x},
\end{equation}
we obtain 
\begin{subequations}\label{axibogtransfli}
\begin{align}
    a_{\xi}^{\regi^{\scriptstyle\dagger}} &= a^{\dagger}_{\xi_-^+} + a_{\xi_-^-} -  b^{\dagger}_{\xi_+^+} - b_{\xi_+^-}, \label{aId} \\
    b_{\xi}^{\regii^{\scriptstyle\dagger}} &= -a^{\dagger}_{\xi_+^+} - a_{\xi_+^-} +  b^{\dagger}_{\xi_-^+} + b_{\xi_-^-} \label{aIId}, 
\end{align}
\end{subequations}
where \(\xi^+_{\pm}(\omega) = \xi_{\pm}(\omega)\) and \(\xi^-_{\pm}(\omega) = \xi_{\pm}^*(-\omega)\) with 
\begin{align}
    \xi_{\pm}(\omega) &= \mathcal E(\omega)\int_{-\infty}^{\infty}dx \theta(\pm x)\xi(x)e^{-i\omega x}. \label{ximxip}
\end{align}
Eqs. \eqref{transffieldinv} can be inverted straightforwardly, and the same procedure leads to
\begin{subequations}\label{axibogtransfl}
\begin{align}
    a_{\xi}^\dagger &= a^{\regi^{\scriptstyle\dagger}}_{\xi_-^+} + a^{\regi}_{\xi_-^-} -  b^{\regii^{\scriptstyle\dagger}}_{\xi_+^+} - b^{\regii}_{\xi_+^-}, \label{axia}\\
    b_{\xi}^\dagger &= - a^{\regi^{\scriptstyle\dagger}}_{\xi_+^+} -a^{\regi}_{\xi_+^-} +  b^{\regii^{\scriptstyle\dagger}}_{\xi_-^+} + b^{\regii}_{\xi_-^-}. \label{axib}
\end{align}
\end{subequations}
Eqs. \eqref{axibogtransfli} (and the inverse \eqref{axibogtransfl}) describe a Bogoliubov transformation involving both creation operators and annihilation operators on the right-hand side.

The Fock space for $t<0$ is based on the vacuum state of regions $\regi$ and $\regii$, $\smallket{0^\regi}\smallket{0^{\regii}}$, defined by $a_\xi^\regi \smallket{0^\regi}\smallket{0^{\regii}} = b_\xi^{\regii} \smallket{0^\regi}\smallket{0^{\regii}} = 0$ for all $\xi(\omega)$. Similarly, for $t>0$ the vacuum state $\smallket{0}\smallket{0}$ for forward- and backward-propagating modes is defined by $a_\xi \smallket{0}\smallket{0} = b_\xi \smallket{0}\smallket{0} = 0$. 

\emph{Truncated photon state.}
Consider an initial single photon in region $\regi$ with frequency spectrum $\xi(\omega)$: 
\begin{equation}\label{initialstate}
    a^{\regi^{\scriptstyle\dagger}}_\xi \ket{0^\regi}\ket{0^{\regii}}.
\end{equation}
The truncated photon state is expressed by formulating \eqref{initialstate} in terms of the ladder operators and vacuum state of forward- and backward-propagating modes, after the reflector has been removed. For $\raisemath{-3pt}{\rule{0pt}{12.6pt}}\smash{a_\xi^{\regi^{\scriptstyle\dagger}}}$ this is just a matter of substituting \eqref{aId}. Expressing $\smallket{0^\regi}\smallket{0^{\regii}}$ in terms of the Fock states of forward- and backward-propagating modes, however, is more involved. In the
\supplementary
% ~\footnote{See Supplemental Material at [] for a detailed derivation of the truncated photon state and an analysis of gradual removal, which includes Refs. \cite{braunstein2005,
% Bogolyubov_trans_general_case,
% C_symmetric_operators,
% vonNeumann1939,
% Ruijsenaars_systems_rel_char_particles,
% lill2022implementingbogoliubovtransformationsshalestinespring,
% shale_linear_1962,
% shale_stinespring_1965,
% chadam_unitarity_1968,
% friedrichs1953mathematical}.
% } 
\nocite{braunstein2005,
Bogolyubov_trans_general_case,
C_symmetric_operators,
vonNeumann1939,
Ruijsenaars_systems_rel_char_particles,
lill2022implementingbogoliubovtransformationsshalestinespring,
shale_linear_1962,
shale_stinespring_1965,
chadam_unitarity_1968,
friedrichs1953mathematical}
we find that
\begin{align}
    & a^{\regi^{\scriptstyle\dagger}}_\xi \ket{0^\regi}\ket{0^{\regii}} = \mathcal N   \lp[ a_{\zeta}^\dagger + b_{\chi}^\dagger \rp] \label{truncphot}\\
    &\cdot \exp\lp[-\frac{1}{2}\int_{-\infty}^\infty dk \int_{-\infty}^\infty dk' \, K(k,k') a^\dagger(k)a^\dagger(k') \rp] \ket{0}\ket{0}. \nonumber
\end{align}
Here, the normalization constant
\begin{equation}
    \mathcal{N} = 
    \big\langle 0 \big| \big\langle 0 \big| 0^{\regi} \big\rangle \big|0^{\regii} \big\rangle
\end{equation}
is the overlap of the two vacua, and the integration kernel $K(k,k')$ and spectra $\zeta$ and $\chi$ can be determined from \eqref{axibogtransfli} 
% (see the \supplementary~\footnotemark[\value{footnote}]). 
(see the \supplementary).

The right-hand side of \eqref{truncphot}, after tracing out the backward-propagating modes, is our truncated photon state $\rho_\xi$.

Computing the exact partial trace over backward-propagating modes is cumbersome. The main difficulty is that the exponential factor in \eqref{truncphot}, acting on $\smallket{0}\smallket{0}$, produces a multimode squeezed vacuum (cf. Eq.~(32) in~\cite{lill2022implementingbogoliubovtransformationsshalestinespring}), which is not written in a form adapted to the forward/backward bipartition. Related squeezed-vacuum structures also appear in relativistic quantum information, where restricting attention to a region of spacetime can make the vacuum appear populated with particles~\cite{shirokov1980,martinetti2003,UnruhEffect_and_applications,ida2013,su2016a,foo2020,onoe2022,camblong2024}. In certain special cases, such as the Unruh effect, the squeezed vacuum admits a convenient form in which the partial trace can be carried out explicitly, yielding a thermal state~\cite{UnruhEffect_and_applications}. Eq. \eqref{truncphot} contains in addition the linear factor \(a_{\zeta}^\dagger+b_{\chi}^\dagger\), which creates a single excitation in a superposition of forward- and backward-propagating modes on top of the squeezed vacuum. It nevertheless follows from \eqref{truncphot} that, after tracing out the backward-propagating modes, the truncated photon state is mixed and generally contains contributions with \(0,1,2,\ldots\) photons, up to infinity.

\emph{Fidelity.} We are interested in the fidelity $F_\xi$ of the truncated photon state $\rho_\xi$ to a forward-propagating single photon state with spectrum $\xi(\omega)$:
\begin{equation}
    F_\xi^2 = \bra{0}a_\xi \rho_\xi a_\xi^\dagger\ket{0}.   \label{eq:fidelity_def}
\end{equation}
We estimate the fidelity for the case where the number of photons produced by the mirror removal is small. More precisely, we require $\expval{n} \ll 1$, where $\expval{n}$ is the expected number of photons in forward-propagating modes if we started with vacuum in both regions $\regi$ and $\regii$. This quantity can be calculated straightforwardly as
\begin{align}
    \expval{n} &= \sum_\xi \bra{0^\regi}\bra{0^{\regii}} a_\xi^\dagger a_\xi \ket{0^\regi}\ket{0^{\regii}} 
    \notag \\
    &= \sum_\xi \lp(\|\xi_-^-\|^2 + \|\xi_+^-\|^2 \rp),\label{enphot}
\end{align}
where the sum over $\xi$ means a sum over a complete set of orthonormal spectra $\xi(\omega)$. The last equality is found by inserting \eqref{axia} and its hermitian conjugate. Similarly, we can use \eqref{axib} to calculate the expected number of photons in backward-propagating modes, which also amounts to $\eqref{enphot}$.

We can express 
\begin{align}
    \ket{0^{\regi}}\ket{0^{\regii}} =  \mathcal N \ket{0}\ket{0} + \sqrt{1-\mathcal N^2} \ket{\text{multiphotons}}.
\end{align}
Since the multiphoton state contributes by at least 2 to the total photon number $2\expval{n}$, we have $1-\mathcal N^2 \leq \expval{n}$. Omitting terms of squared norm $\mathcal O(\expval{n})$, and substituting \eqref{aId}, we find
\begin{equation}
    a^{\regi^{\scriptstyle\dagger}}_\xi \ket{0^\regi}\ket{0^{\regii}} = \lp(a^{\dagger}_{\xi_-^+} -  b^{\dagger}_{\xi_+^+} \rp)  \ket{0}\ket{0}.
\end{equation}
Tracing out the backward-propagating modes,
the fidelity \eqref{eq:fidelity_def} is thus
\begin{equation}
    F_{\xi} = |\bra{0}a_\xi a_{\xi^+_-}^\dagger \ket{0}| = \lp| \langle\xi,\xi^+_-\rangle \rp|. \label{eq:def:F_xi}
\end{equation}
Using the Plancherel theorem (\cite[Ch. 11]{mcdonald_weiss2012}) gives
\begin{align}
    \langle \xi,\xi^+_-\rangle &= \int_0^\infty d\omega \xi^*(\omega)\xi_-(\omega) \label{olxixipm} = \int_{-\infty}^\infty d\omega \mathcal E(\omega)\xi^*(\omega) \frac{\xi_-(\omega)}{\mathcal E(\omega)} \nonumber\\ 
    &= \int_{-\infty}^\infty dx \theta(-x) \tilde\xi^*(x) \xi(x),
\end{align}
where
\begin{equation}
    \tilde\xi(x) = \int_0^\infty d\omega \mathcal E(\omega)\xi(\omega)e^{i\omega x}.
\end{equation}
Thus, the fidelity in the special case $\expval{n} \ll 1$ becomes
\begin{equation}\label{Fxifid}
    F_\xi = \lp| \int_{-\infty}^\infty dx \theta(-x) \tilde\xi^*(x) \xi(x) \rp|.
%    F_\xi = \lp(1 - \expval{n} \rp) \lp| \int_{-\infty}^0 dx \tilde\xi^*(x) \xi(x) \rp|.
\end{equation}
This is a quite intuitive result: Both $\xi(x)$ and $\tilde\xi(x)$ can be seen as ``photon wave functions'' for $t=0$, and the fidelity is given by the ``fraction of the photon'' located at $x<0$.

\emph{Gradual removal.} Unfortunately, for instantaneous mirror removal it follows from \eqref{ximxip} and \eqref{enphot} that $\expval{n}=\infty$. This is because there are modes $\xi(\omega)$ such that $\xi(x)\neq 0$ for $x=0$; then the Heaviside function leads to a Fourier transform with asymptotic behavior $\propto 1/\omega$ for large $\omega$. Together with $\mathcal E(\omega)\propto \sqrt{\omega}$, the norms diverge logarithmically. Physically, the instantaneous mirror removal produces an infinite number of photons \cite{cirone1999,braunstein2005}. To bound $\expval{n}$, we must therefore consider a smoother process. 

% In the \supplementary~\footnotemark[\value{footnote}], 
In the \supplementary, 
we demonstrate that we can obtain a finite $\expval{n}$ by considering a gradual removal, from $t=-T$ to $t=0$, of an infinitely thin mirror. Eq. \eqref{truncphot} is still valid, with a well-defined $\mathcal N$. Concretely, we consider the dielectric
\(\epsilon(x,t) = 1 + \kappa(t)\delta(x)\), where \(\kappa(t)\) is the integrated permittivity of a thin dielectric slab~\cite{braunstein2005}.  We implement the removal of the mirror by letting \(\kappa(t)\) be a constant value \(\kappa_0\) for \(t\leq-T\), and a monotonically decreasing quadratic function for \(-T<t\leq 0\). This leads to
\begin{align}
    \expval{n} \leq \frac{\kappa_0}{4T} + \frac{\kappa_0^2}{16T^2}. \label{bound_expval_n}
\end{align}
The parameter \(\kappa_0\) is directly related to the initial transmissivity $|\mathcal T(\omega)|^2$ of the mirror:
\begin{equation}
    |\mathcal T(\omega)|^2 = \frac{1}{1+(\omega\kappa_0)^2/4}.
\end{equation} 
To ensure a negligible transmission coefficient for the incident photon prior to removing the mirror, we require \(\omega_0\kappa_0\gg 1\), where \(\omega_0\) is the central frequency of the photon. 
To obtain \(\expval{n}\lesssim 1\), \eqref{bound_expval_n} then requires
\begin{align}
    1/\omega_0\ll \kappa_0 \lesssim T.
\end{align}
At the same time, $T$ should be small enough such that the mirror truncates the photon without changing its shape otherwise, i.e., $T$ must be smaller than the temporal variation scale of the photon pulse envelope.

As a concrete example, for an optical photon with \(\omega_0/2\pi = 10^{15}\)~Hz, requiring $|\mathcal{T}(\omega_0)|^2=10^{-4}$ allows the parameter \(T\) to be as small as \(10^{-14}\)~s before \(\expval{n}\) becomes comparable to~1. 

From now on, we assume that the removal process is gradual. After the mirror has been turned off, there will be a bounded \emph{transition region} $t<x<t+T$ where the forward-propagating field depends on the gradual removal process.

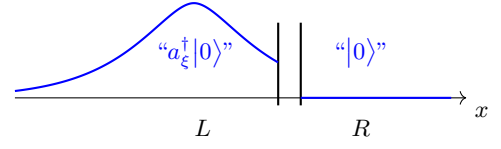
\begin{figure}[t]
\begin{center}
\begin{tikzpicture}[scale=1.0]

% Horizontal axis
\draw[->] (-3.5,0) -- (2.5,0) node[below right] {$x$};

% Transition region:
\draw[thick] (0,-0.1) -- (0,1);
\draw[thick] (0.3,-0.1) -- (0.3,1);

% Incoming wave (with exp(-x) photon tails)
\draw[blue,thick,smooth] plot[domain=-5:0,samples=200,scale=0.7] (\x,{1.8*exp(-(\x+1.6)^2/(1+abs(\x+1.6)))});
\draw[thick,blue] (0.3,0) -- (2.3,0);

% Labels, \ket{1} and \ket{0}
\node[blue] at (-1.1,0.6) {\(\text{``}a_\xi^\dagger\ket{0}\text{''}\)};
\node[blue] at (1.1,0.6) {\(\text{``}\ket{0}\text{''}\)};

\node[black] at (-1.0,-0.4) {$L$};
\node[black] at (1.1,-0.4) {$R$};

\end{tikzpicture}
\caption{Considering forward-propagating modes, the truncated photon state is locally equivalent to a single photon $a_\xi^\dagger\smallket{0}$ to the left of the transition region, and vacuum $\smallket{0}$ to the right. Equivalence means that all local observables give the same measurement statistics. In the main text we consider observables $L$ and $R$ on the left-hand side and right-hand side of the transition region, respectively.}
\label{fig:equiv} 
\end{center}
\end{figure}

\emph{Local equivalence.} Outside the transition region the truncated photon state has a remarkable property, which we believe is of interest for theoretical quantum optics and quantum field theory. Recall that the state \eqref{truncphot}, after tracing out backward-propagating modes, is not particularly nice; it contains superpositions and mixtures of terms with unbounded photon numbers. Yet, it looks exactly like the vacuum state to the right of the transition region, and exactly like a single-photon state to the left of the transition region (Fig. \ref{fig:equiv}). The truncated photon state is thus an example of a very complicated state that produces the exact same measurement statistics as very simple states, as long as one is interested only in local observables to the left or right of the transition region.

To prove the local equivalence of the truncated photon state and a single photon state for $x<t$, we first establish the following definition \cite{knight1961,haag1996,gulla2024}: a local observable $L$ is an operator expressible as a superposition of products of fields evaluated in the measurement region. Here we are interested in the forward-propagating field for $t>0$, meaning that local observables are expressible from $E_a(t-x)$ evaluated at the time and place of the measurement. 

We consider the expectation value of an observable $L$ local to the region $x<t$. From \eqref{transffieldinv} (which still applies outside of the transition region) we have ${E_a(t-x) = E^{\regi}(t-x)}$ for $x<t$, so $L$ must also be expressible from $E^{\regi}(t-x)$ there. Hence,
\begin{align}\label{Lequiv}
    &\bra{0^\regi}\bra{0^{\regii}} a^{\regi}_\xi L a^{\regi^{\scriptstyle\dagger}}_\xi \ket{0^\regi}\ket{0^{\regii}} \\
    &= \bra{0^\regi} a^{\regi}_\xi L a^{\regi^{\scriptstyle\dagger}}_\xi \ket{0^\regi} = 
    \bra{0} a_\xi L a^{\dagger}_\xi \ket{0}. \nonumber
\end{align} 
The last equality is subtle; it is obtained by thinking of $L$ as being expressed from $E^{\regi}(t-x)$ in $\smallbra{0^\regi} a^{\regi}_\xi L a^{\regi^{\scriptstyle\dagger}}_\xi \smallket{0^\regi}$, while in $\smallbra{0} a_\xi L a^{\dagger}_\xi \smallket{0}$ it is expressed from $E_a(t-x)$. Since the formal relation between $\smallket{0^\regi}$, $a_\xi^\regi$, and $E^{\regi}$ is exactly the same as the relation between $\smallket{0}$, $a_\xi$, and $E_a$, the result from calculating the expectation values will be the same.

To prove the local equivalence of the truncated photon state and vacuum for $x>t+T$, we consider an observable $R$ local to $x>t+T$, i.e., an operator expressible from ${E_a(t-x)}$ for $x>t+T$. Eq. \eqref{transffieldinv} implies that $E_a(t-x)=-E^{\regii}(t-x)$ there. A similar calculation then gives
\begin{align}\label{Requiv}
    \bra{0^\regi}\bra{0^{\regii}} a^{\regi}_\xi R a^{\regi^{\scriptstyle\dagger}}_\xi \ket{0^\regi}\ket{0^{\regii}} 
    = \bra{0^{\regii}} R \ket{0^{\regii}} = \bra{0} R \ket{0}.
\end{align}

Remarkably, by letting the transition region be small, we obtain a state with large expected number of photons, that looks like vacuum or a single photon everywhere except in a narrow transition region. This sounds like a paradox, since \eqref{Lequiv} and \eqref{Requiv} imply that the energy density is equal to that of a single photon or vacuum, except in the transition region.

The resolution comes from the realization that the produced photonic state leads to a high energy density inside the transition region. 
To calculate the energy density excluding the contribution from the vacuum fluctuations, one may normal order the squared fields in the energy density expression (see, e.g., \cite{bialynicki-birula1998}). Since the quantity of interest is the energy density \emph{after} the mirror has been removed, normal ordering must be in terms of the forward-propagating ladder operators (for $t>0$), not the ladder operators of regions $\regi$ and $\regii$. For a small transition region, the difference between these two normal orderings is substantial, as the Bogoliubov transformation mixes the creation and annihilation operators strongly. 
Physically, we may say that the mirror removal converts vacuum energy to real photon energy in the forward- and backward-propagating modes.

The local equivalence of the truncated photon state and vacuum for $x>t+T$ means that the state is \emph{strictly localized} to $x<t+T$ \cite{knight1961,licht1963}\footnote{Strictly speaking, one should say ``arbitrarily well localized'', since for a finite $\kappa_0$ the mirror transmissivity is not exactly zero. Similar reservations apply to the local equivalence statements above.}. This is also evident from causality. From Knight's theorem \cite{knight1961} it is therefore not surprising that the state \eqref{truncphot} involves arbitrarily high photon numbers. This is the case even if the reflector is turned off slowly. We also note that the fidelity \eqref{Fxifid} is compatible with the upper and lower fidelity bounds for strictly localized states \cite{gulla2023}.

Even though the truncated photon state is strictly localized to $x<t+T$, the state \eqref{truncphot} \emph{before} tracing out the backward-propagating modes is \emph{not}, considering both the forward and backward-propagating parts of the field. Indeed, one can show explicitly that \eqref{truncphot} is instead locally equivalent to the vacuum state $\smallket{0^{\regii}}$ for $x>t+T$, which in turn is not locally equivalent to $\smallket{0}\smallket{0}$. Intuitively, this is because $\smallket{0^{\regii}}$ is the vacuum state associated to a field where the  forward- and backward-propagating parts are correlated (due to the mirror), while $\smallket{0}\smallket{0}$ is the vacuum state associated to a field with uncorrelated forward- and backward-propagating parts. In other words, if the state of the system is $\smallket{0^{\regi}}\smallket{0^{\regii}}$ one measures interference between forward and backward-propagating vacuum fluctuations. Due to causality, the situation does not change for $x>t+T$ even though the mirror is gone. Thus for $x>t+T$ the state $\smallket{0^{\regi}}\smallket{0^{\regii}}$ does not look like $\smallket{0}\smallket{0}$, for which one measures no such interference.

Having discussed how the truncated photon demonstrates the idea of local equivalence and strict localization, it is natural to mention finally that it is a special case of states perturbed by classical, localized interactions. While photonic states can be produced on demand by simple components in the lab, and therefore by causality can be strictly localized, massive particles such as electrons are usually present to begin with. In the lab it is usually only a \emph{perturbation} of the particle state that is provided by the trigger. By causality, the perturbation, not the particle state itself, is localized to the light cone of the trigger. The truncated photon is a natural example that may help study the set of such causally perturbed states.

\appendix

\nocite{braunstein2005,
Bogolyubov_trans_general_case,
C_symmetric_operators,
vonNeumann1939,
Ruijsenaars_systems_rel_char_particles,
lill2022implementingbogoliubovtransformationsshalestinespring,
shale_linear_1962,
shale_stinespring_1965,
chadam_unitarity_1968,
friedrichs1953mathematical}

\section*{\supplementary}
In this supplemental material, we derive the form of the truncated photon state (before tracing out the backward-propagating modes) and analyze a gradual removal of the reflector. This material is divided into two sections, Appendix \ref{app:bogoliubov} and Appendix \ref{app:gradual_removal}.

In the first section, we begin by treating some of the analytic properties of Bogoliubov transformations. This involves stating the so-called Shale-Stinespring condition, which, when satisfied, guarantees that the new vacuum state exists in the Fock space. We also derive a lower bound for the fidelity of the one-particle content of the truncated photon state before the backward-propagating modes have been traced out. A critical parameter for these results is the expected photon number in an initial vacuum state.

The analysis of a gradual removal of the reflector in Appendix \ref{app:gradual_removal} is motivated by the divergent photon number found for the instantaneous removal of the reflector in the main text. We find a finite photon number of the vacuum state, with an upper bound in terms of the removal time and initial reflectivity of the reflector. 

\section{The Bogoliubov transformation and truncated photon}    \label{app:bogoliubov}
This section provides a formal description of both the Bogoliubov transformation resulting from the removal of the reflector and properties of the truncated photon state. We begin by presenting some key aspects of the Bogoliubov transformation, and, in particular, its associated new vacuum state. This involves introducing some general notation to parameterize Bogoliubov transformations in a convenient way. Using this notation, we then proceed with deriving the form of truncated photon state before tracing out backward-propagating modes, and finally calculate a one-particle fidelity in terms of the expected photon number of the vacuum. 

\subsection{Bogoliubov transformation}

In the following, we write
\begin{align}
    a_{f} = \int dk f^*(k) a(k), \quad a_{f}^{\dag} = \int dk f(k) a^{\dag}(k),
\end{align}
where $f\in \mathcal H \equiv L^2(-\infty,\infty)$, the one-particle Hilbert space. Here $a(k)$ is the usual annihilation operator as a function of one-dimensional wavenumber $k$. We let the notation $f^{+}$ and $f^{-}$ represent functions supported on $(0,\infty)$ and $(-\infty,0)$, respectively, corresponding to forward- and backward-propagating modes, respectively.

A general Bogoliubov transformation can be characterized by two operators, \(T_1\) and \(T_2\) on \(\mathcal{H}\), which describe how the original set of ladder operators \(\smash{a_{f}}, \smash{a^{\dag}_{f}}\) transforms to a new set \(\smash{a'_{f}}, \smash{a'^{\dag}_{f}}\). That is,
\begin{align}
    a'_{f} = a_{T_1f} + a_{T_2Cf}^{\dag}, \quad f\in\mathcal{H}.    \label{generalBogo}
\end{align}
% Here, $C$ is the generalized complex conjugation operator, defined by $\langle Cf,g\rangle = \langle C^{\dag}g,f\rangle$, with ${C=C^{-1}=C^{\dag}}$.
Here, \(C\) is the complex conjugation operator, defined by $CAf = (Af)^*$, where $A$ is a general operator. Its adjoint is understood as $\langle Cf,g\rangle = \langle C^{\dag}g,f\rangle $, for ${f,g\in\mathcal{H}}$,  and ${C=C^{-1}=C^{\dag}}$. 
Note that the expression $T_2Cf$ is equal to $T_2 f^*$, which is different from $CT_2f$. 
The parameterization of Bogoliubov transformations by use of the two operators \(T_1\) and \(T_2C\), or $CT_2$, is common in the mathematical literature on Bogoliubov transformations~\footnote{See, for instance, Refs. \cite{Ruijsenaars_systems_rel_char_particles, Bogolyubov_trans_general_case}, or Appendix B in \cite{lill2022implementingbogoliubovtransformationsshalestinespring}.}.

A Bogoliubov transformation introduces a new vacuum state. This is seen from \eqref{generalBogo}, since if \(a_{f}\) annihilates a state for all $f\in\mathcal H$, \(a'_{f}\) does not when \(T_2\) is nonzero. Requiring the operator \(T_2\) to be Hilbert-Schmidt (H.S), i.e. that it admits a finite H.S. norm, is required for the implementability of a new vacuum state on the underlying Fock space. This is the so-called Shale-Stinespring condition \footnote{See Refs. \cite{shale_linear_1962,shale_stinespring_1965,chadam_unitarity_1968}, although as mentioned in~\cite{Bogolyubov_trans_general_case}, the original ``semirigorous'' work was done already in \cite{friedrichs1953mathematical}. It can be possible to implement a new vacuum state on an extended Fock space in scenarios where the Shale-Stinespring condition is not met \cite{lill2022implementingbogoliubovtransformationsshalestinespring}}. The H.S. norm is given by
\begin{align}
    \|T_2\|_{HS}^2 = \sum_{n} \|T_2e_n\|^2,
\end{align}
where \(e_n\) is some orthonormal basis for $\mathcal H$. 

In the scenario described in the main text, with the removal of a reflector located at $x=0$, we obtain a Bogoliubov transformation from the temporal continuity of the fields at $t=0$. To describe this resulting Bogoliubov transformation in the notation just introduced, we recall the field transformation at $t=0$ (see main text):
\begin{subequations}    \label{transffieldinv_appendix}
    \begin{align}
        E^{\text{I}}(x) &= \theta(x)E_a(x) - \theta(-x)E_b(x), \label{fieldtransI} \\
        E^{\text{II}}(x) &= -\theta(-x)E_a(x) + \theta(x)E_b(x). \label{fieldtransII}
    \end{align}
\end{subequations}
Here, superscript I and II refer to the region $x<0$ and $x>0$, respectively, while $E_a$ and $E_b$ denote the forward- ($a$) and backward-propagating ($b$) electric fields for $t>0$.

In Appendix \ref{app:gradual_removal} we will consider a gradual removal of the reflector and end up with a transformation which is different from  \eqref{transffieldinv_appendix}. However, the following analysis, and in fact, the remainder of this Appendix will hold for any general transformation where $E^{\regi}$ and $E^{\regii}$ transform separately into some combination of $E_a$ and $E_b$. In particular, for the transformation we find in Appendix \ref{app:gradual_removal}.

Taking the Fourier transform of \eqref{transffieldinv_appendix} and integrating \eqref{fieldtransI} with \(f^{+}(k)\) and \eqref{fieldtransII} with \(f^{-}(k)\) leads to the following relations:
\begin{subequations}    \label{positive_negative_modes_relation}
    \begin{align}
        a_{f^+}^{\text{I}} &= a_{T_1^{\text{I}}f^+} + a_{T_2^{\text{I}}Cf^+}^{\dag}, \label{Bogoliubov_trans_firstI} \\
        a_{f^-}^{\text{II}} &= a_{T_1^{\text{II}}f^-} + a_{T_2^{\text{II}}Cf^-}^{\dag}, \label{Bogoliubov_trans_firstII}
    \end{align}
\end{subequations}
for some operators \(T_{1}^{\text{I}}\), \(T_{2}^{\text{I}}\) and \(T_{1}^{\text{II}}\), \(T_{2}^{\text{II}}\) which act exclusively on forward- and backward-propagating modes, respectively.
By letting the operators $T_1$ and $T_2$ in \eqref{generalBogo} be given by
\begin{subequations}
    \begin{align}
        T_{1} &= T_1^{\text{I}} + T_1^{\text{II}}, \\
        T_{2} &= T_2^{\text{I}} + T_2^{\text{II}},
    \end{align}
\end{subequations}
% and 
we obtain 
\begin{align}
    a'_{f} = a^{\text{I}}_{f^+} + a^{\text{II}}_{f^-},  
    \label{ourBogo}
\end{align}
according to \eqref{positive_negative_modes_relation} and \eqref{generalBogo}. 

Let \(\mathcal{U}\) be the \emph{unitary} operator implementing the Bogoliubov transformation \eqref{generalBogo}:
\begin{align}
    a'_{f} = \mathcal{U}a_{f}\mathcal{U}^{\dag}
    % = a_{T_1f} + a_{T_2Cf}^{\dag}. 
    \label{mathcalUBogo}
\end{align}
Its corresponding inverse Bogoliubov transformation is given by \cite{Bogolyubov_trans_general_case}
\begin{align}
    \mathcal{U}^{\dag}a'_{f}\mathcal{U} = a'_{T_{1}^{\dag}f} - a'^{\dag}_{CT_{2}^{\dag}f}. \label{inverseBogo}
\end{align}  

The vacuum $\smallket{0}\smallket{0}$ is defined as the state annihilated by all operators $a_f$. The state is written with two $\smallket{0}$'s to emphasize that it is a tensor product of vacuum for forward- and backward-propagating modes. Since $\mathcal{U}$ is unitary, \eqref{mathcalUBogo} implies that the state
\begin{align}
    \ket{0^{\text{\text{I}}}} \ket{0^{\text{\text{II}}}} \equiv \mathcal{U}\ket{0}\ket{0} \label{ourVacuum}
\end{align}
is annihilated by all $a_f'$. Therefore, \eqref{ourVacuum} describes the vacuum state of the two regions I and II (for $t<0$).

Requiring \(\mathcal{U}\) to be unitary corresponds to demanding that the transformed ladder operators satisfy the canonical commutation relations, and leads to a set of four constraint equations on the operators \(T_1\) and \(T_2\) \footnote{The canonical commutators for a general ladder operator $a_f$ are: ${[a_{f}, a^{\dag}_{g}] = \langle f, g\rangle}$ and ${[a_{f}, a_{g}] = [a^{\dag}_{f}, a^{\dag}_{g}] = 0}$, where $f,g\in\mathcal{H}$. Ensuring \eqref{mathcalUBogo} and \eqref{inverseBogo} satisfy these leads to Eqs. \eqref{constraint1} and \eqref{constraint2}, and Eqs. \eqref{constraint3} and \eqref{constraint4}, respectively.}. In the parameterization we use here, these are \footnote{The constraints on $T_1$ and $T_2$ in \eqref{constraints} may look different for different ways of parameterizing Bogoliubov transformations, see for instance Appendix B in \cite{lill2022implementingbogoliubovtransformationsshalestinespring}.}:
\begin{subequations}    \label{constraints}
\begin{align}
    T_1^{\dag}T_1 - CT_2^{\dag}T_2C &= 1,\label{constraint1}  
    \\ 
    T_1^{\dag}T_2 - CT_2^{\dag}T_1C &= 0, \label{constraint2}\\
    T_{1}T_{1}^{\dag} - T_{2}T_{2}^{\dag} &= 1, \label{constraint3} \\
    T_1CT_2^{\dag}C - T_2CT_{1}^{\dag}C &= 0.   \label{constraint4}
\end{align}
\end{subequations}
The constraints \eqref{constraint1} and \eqref{constraint2} ensure that $\mathcal{U}^{\dag}\mathcal{U}=1$, while  \eqref{constraint3} and \eqref{constraint4} ensure that $\mathcal{U}\mathcal{U}^{\dag}=1$. 

Lastly, we can relate the Shale-Stinespring condition to the expected photon number of the vacuum state:
\begin{align}
    \langle n_{\text{tot}}\rangle &\equiv \sum_{n}\bra{0^{\text{I}}}\bra{0^{\text{II}}}a^{\dag}_{e_n}a_{e_n}\ket{0^{\text{I}}}\ket{0^{\text{II}}} 
    \notag \\
    &= \sum_{n} \bra{0}\bra{0}\mathcal{U}^{\dag}a^{\dag}_{e_n}a_{e_n}\mathcal{U}\ket{0}\ket{0} = \|T_{2}\|_{\text{H.S.}}^{2}.    \label{expectPhoton_T_2}
\end{align}
The last equality follows from the expression for the inverse Bogoliubov transformation \eqref{inverseBogo} and the fact that $\|T_{2}^{\dag}\|_{\text{H.S.}}^{2} = \|T_{2}\|_{\text{H.S.}}^{2}$. Eq. \eqref{expectPhoton_T_2} means that a finite photon number is required for the Shale-Stinespring condition to be satisfied, in other words, for \eqref{ourVacuum} to be well-defined.
% Note that a consequence of \eqref{constraints} is that \(\text{ran}(T_1) = \mathcal{H}\) and \(\|T_1\|\geq 1\) \footnote{Eq. \eqref{constraint3} implies that \(\text{ker}(T_{1}^{\dag})=0\) (assume otherwise leads to a contradiction), which in turn means that \(\mathcal{H} = \text{ker}(T_{1}^{\dag})^{\bot} = \overline{\text{ran}(T_1)}\), where \(\bot\) and the overline denotes the orthogonal complement and closure, respectively. Thus, \(\overline{\text{ran}(T_1)} = \text{ran}(T_1)\) since \(T_1\) is bounded from below, according to \eqref{constraint1}. The fact that \(\|T_1\|\geq 1\) follows simply from \eqref{constraint1}.}. In turn, this can be used to show that \(\|T_{1}^{-1}\|\leq 1\) \footnote{The norm of \(T_{1}^{-1}\) obeys \(\|T_{1}^{-1}\|\leq 1\), which is true because \(\|T_{1}^{-1}\| = \sup_{g\neq 0} \|T_{1}^{-1}g\|/\|g\|=\sup_{f\neq 0}\|f\|/\|T_1f\|\leq 1\), which follows from that \(\text{ran}(T_{1}) = \mathcal{H}\) and that \(\|T_1\|\geq 1\).}. \emph{Det siste resultatet må sjekkes.}

% To establish the unitarity of \(\mathcal{U}\), we will show that our field transformations preserve the commutation relations. This implies that the ladder operators also preserve their commutation relations, which in turn establishes that \(\mathcal{U}\) is unitary. 

\subsection{Vacuum state}   \label{subsec:new_vacuum_sate}
When the mirror is present ($t<0$), the vacuum state is $\ket{0^{\text{\text{I}}}} \ket{0^{\text{\text{II}}}}$. After the mirror has been removed, it is useful to express this state in terms of the vacuum and ladder operators of forward- and backward-propagating modes. This is done with \eqref{ourVacuum} and, e.g., Lemma 5.1 and Eq. (5.3) in~\cite{Bogolyubov_trans_general_case} (compared to the notation in \cite{Bogolyubov_trans_general_case}, our operators \(T_1\) and \(T_{2}\) correspond to \(U_{++}\) and \(-CU_{-+}C\), respectively):
\begin{align}
    & \mathcal{U}\ket{0}\ket{0} = 
    \mathcal{N} \label{newVacuumState} \\
    & \cdot\exp\left[-\frac{1}{2}\int_{-\infty}^{\infty}dk \int_{-\infty}^{\infty}dk' K(k,k')a^{\dag}(k)a^{\dag}(k')\right]\ket{0}\ket{0}, \nonumber
\end{align}
where $K(k,k')$ is the integral kernel of 
\begin{align}
    K = T_2 CT_1^{-1}C.   \label{T_U_def}
\end{align}
The normalization $\mathcal{N}$ in \eqref{newVacuumState} is given by 
 \begin{align}
    \mathcal{N} = \bra{0}\bra{0} \mathcal U \ket{0}\ket{0} = \prod_{n}\left(1 - \lambda_n^2\right)^{\frac{1}{4}},   \label{N_product}
\end{align}
where $\lambda_n$ are the eigenvalues of the operator \(|K|\) (see (4.10) in~\cite{Bogolyubov_trans_general_case}). These eigenvalues exist because the operator $K$ obeys ${K^{\dag}=CKC}$, as can be seen from \eqref{constraint2}. Such operators are called $C$-symmetric and admit a singular-value decomposition (see Theorem 3 in~\cite{C_symmetric_operators}). 

%  The determinant in \eqref{newVacuumState} serves as a normalization factor for the new vacuum state, and will play a crucial role in the fidelities for the truncated photon. Thus, for notational ease, we define 
% \begin{align}
%     \mathcal{N} = \det\left(1 - T_{\mathcal{U}}^{\dag}T_{\mathcal{U}}\right)^{\frac{1}{4}}. \label{normalizationFactor}
% \end{align}
% By identifying the eigenvalues\footnote{See (4.10) in \cite{Bogolyubov_trans_general_case}} \(\lambda_n\) of the operator \(|T_{\mathcal{U}}|\), the normalization factor (the determinant) in \eqref{normalizationFactor} can be computed as 

The normalization $\mathcal{N}$ is well-defined if both \(\mathcal{U}\) is unitary and the Shale-Stinespring condition is satisfied. Unitarity leads to \(\lambda_n<1\) \cite[(4.9)]{Bogolyubov_trans_general_case}, in which case the infinite product in \eqref{N_product} is nonzero if \(\sum_n \lambda_n^2 < \infty\) (see \cite[Lemma 2.4.1]{vonNeumann1939}). 
This sum converges because the operator \(K\) is H.S, which follows from the Shale-Stinespring condition \footnote{\(K\) is a composition of a H.S. operator (\(T_2\)) and a bounded operator (\(CT_1^{-1}C\)), thus it is itself H.S.}. Indeed, we have that
\begin{align}
    \|K\|_{\text{H.S.}}^{2} &=
    \sum_{n} \lambda_n^2,  \label{T_U_hs_norm}
\end{align}
which is found from the definition of the H.S. norm and the fact that $|K|$ has an orthonormal basis.

\subsection{Truncated photon state}    \label{sec:TruncatedPhotonState}
In the main text we consider a single photon state propagating in from the left, $a^{\text{\text{I}}\dag}_{\xi} \ket{0^{\regi}} \ket{0^{\regii}}$. 
When the reflector has been removed ($t>0$), and after the backward-propagating modes have been traced out, we refer to this state as the truncated photon state. Using relation \eqref{ourVacuum} between the two vacua, the Bogoliubov transformation \eqref{mathcalUBogo}, and (4.47) in \cite{Bogolyubov_trans_general_case}, we get that
\begin{align}
   a^{\text{\text{I}}\dag}_{\xi} \ket{0^{\text{\text{I}}}} \ket{0^{\text{\text{II}}}} &= \left(a_{T_1\xi}^{\dag} + a_{T_2C\xi}\right) \mathcal{U}\ket{0}\ket{0}
    \notag \\
    &= \left(a^{\dag}_{T_1\xi} - a^{\dag}_{KCT_2C\xi}\right)\mathcal{U}\ket{0}\ket{0} \notag\\
    &= a^{\dag}_{\zeta}\mathcal{U}\ket{0}\ket{0},
\end{align}
where we have defined 
\begin{align}
    \zeta = (T_1-KCT_2C)\xi = T_{1}^{\dag-1}\xi. \label{eq:zeta_definition} 
\end{align}
Together with \eqref{newVacuumState} we finally obtain
\begin{align}
    & a^{\regi^{\dag}}_{\xi} \ket{0^{\regi}} \ket{0^{\regii}} 
    = \mathcal{N} a_{\zeta}^{\dag} \label{eq:initial_state_written_out}\\ & \cdot\exp\left[-\frac{1}{2}\int_{-\infty}^{\infty}dk \int_{-\infty}^{\infty}dk' K(k,k')a^{\dag}(k)a^{\dag}(k')\right] \ket{0}\ket{0}. \nonumber
\end{align}
Here $\zeta(k)$ is supported for both positive and negative $k$. Therefore \eqref{eq:initial_state_written_out} is the same as the expression in the main article for the truncated photon state before the backward-propagating modes are traced out. In the main article the part of $\zeta$ supported for negative $k$ has been taken out and denoted $\chi$.

\subsection{One-particle fidelity}
The one-particle content of \eqref{eq:initial_state_written_out} can be measured by the fidelity
\begin{align}
    F_{1}^2 \equiv |\bra{0^{\regi} }\bra{0^{\regii}}  a_{\xi}^{\regi}P_1a^{\regi^{\dag}}_{\xi} \ket{0^{\regi}} \ket{0^{\regii} }| = \mathcal{N}^2\|\zeta\|^2,
\end{align}
where $P_1$ is the projector onto the one-particle sector. Note that we here consider forward- \emph{and} backward-propagating modes. That is, even if $F_1$ is close to unity, the truncated photon state (which results after tracing out the backward-propagating modes) may be quite different from a single photon. This is because the leading tail of the incident state has been reflected. 

We will now show that the fidelity $F_1$ can be bounded from below in terms of $\langle n_{\text{tot}}\rangle$. Note first that a consequence of \eqref{constraint1} is that $\|T_1f\|\geq \|f\|$, so $T_1$ is  injective~\footnote{
% It is possible to show that $T_1$ is also surjective using \eqref{constraint1} and \eqref{constraint3}. Concretely, \eqref{constraint3} implies that $\text{ker}(T_1^{\dag})= \{0\}$, while \eqref{constraint1} that the range of $T_{1}$ is closed. Hence, $\mathcal H=\text{ran}(T_1)$.
It is also possible to show that $T_1$ is surjective. Eq. \eqref{constraint3} implies that $\text{ker}(T_1^{\dag})= \{0\}$, while \eqref{constraint1} that the range of $T_{1}$ is closed. Hence, $\text{ran}(T_1)=\mathcal H$.
}. For $f\in\text{ran}(T_1)$, let $g$ be defined by $T_1g=f$. Then $\|T_1^{-1}f\| = \|g\|\leq \|T_{1}g\| = \|f\|$, and so
\begin{align}
    \|T_{1}^{-1}\| 
   \leq 1.  \label{T1_inverse_bound}
\end{align}
Using \eqref{T_U_def}, \eqref{expectPhoton_T_2}, and \eqref{T1_inverse_bound} we find $\|K\|_{\text{H.S.}}^{2}\leq \langle n_{\text{tot}} \rangle$. If we then assume $\langle n_{\text{tot}}\rangle \leq 1$, we obtain the following lower bound for the normalization factor
\begin{align}
    \mathcal{N} \geq \left(1 - \langle n_{\text{tot}}\rangle\right)^{\frac{1}{4}}. \label{lower_bound_N_photon_number}
\end{align}
This can be found by writing the normalization factor as 
$\mathcal{N}=\exp(\frac{1}{4}\sum_n\ln(1-\lambda_n^2))$, expanding the logarithm and using $\|K\|_{\text{H.S.}}^{2}\leq 1$ together with \eqref{T_U_hs_norm}.

Next, considering the expression for $\zeta$ in \eqref{eq:zeta_definition}, and bounding $T_{1}^{\dag}$ from above using \eqref{constraint3}, the general property that $\|\cdot\| \leq \|\cdot\|_{\text{H.S.}}$ and \eqref{expectPhoton_T_2}, we find that 
\begin{align}
    \|\zeta\|^2 \geq \frac{\|\xi\|^2}{\|T_{1}^{\dag}\|^2}\geq \frac{1}{1 + \langle n_{\text{tot}}\rangle}. \label{eq:zeta_lower_bound}
\end{align}

Combining \eqref{eq:zeta_lower_bound} and \eqref{lower_bound_N_photon_number} results in 
\begin{align}
    F_1^2 \geq \frac{(1 - \langle n_{\text{tot}}\rangle)^{\frac{1}{2}}}{1 + \langle n_{\text{tot}}\rangle}, \label{eq:F_1_bound}
\end{align}
which holds for $\langle n_{\text{tot}}\rangle \leq 1$.

The expression \eqref{eq:F_1_bound} means that $F_1 \approx 1$ can be achieved by ensuring a sufficiently small $\langle n_{\text{tot}}\rangle$. In the next section, we derive a bound for $\langle n_{\text{tot}}\rangle$ and identify under what circumstances this quantity is in fact sufficiently small.

\section{Gradual removal} \label{app:gradual_removal}
Having described the general form of the Bogoliubov transformation, we consider the special case of a mirror removal in more detail. In the main text we mainly discussed an instantaneous removal, while we will now treat the more general case of gradual removal. We start with the general form of the linear transformation, and then specialize to a time-dependent electromagnetic reflector.

\subsection{General form} 
 We have a linear transformation of the fields:
\begin{subequations} \label{transffieldgen}
\begin{align}
    E_a(x) &= \int dy g(x,y)E^{\regi}(y) + \int dy h(x,y)E^{\regii}(y), \label{transfEcfieldgen}\\
    E_b(x) &= \int dy h(x,y)E^{\regi}(y) + \int dy g(x,y)E^{\regii}(y), \label{transfEdfieldgen}
\end{align} 
\end{subequations}
for some real functions $g(x,y)$ and $h(x,y)$. Since the mirror is effective only for negative time, the general forms of $g(x,y)$ and $h(x,y)$ are
\begin{subequations}\label{gisak}
\begin{align}
g(x,y) &= \theta(x)\delta(x-y) + \theta(-x) f(x,y), \label{gIisak}\\ 
h(x,y) &= -\theta(-x)\delta(x-y) + \theta(-x)f(x,y). \label{gIIisak}
\end{align}
\end{subequations}
Here $f(x,y)$ is some function that is taken to be the same in \eqref{gIisak} and \eqref{gIIisak} since the mirror is assumed symmetric about $x=0$. This will become clearer in the next section, where we implement the mirror as an infinitely thin, dielectric slab of infinite permittivity, and deduce the corresponding $f(x,y)$. Also note that $f(x,y)=0$ for $y>x$ due to causality, and we may also take $f(x,y)=0$ for $x>0$.

Taking the Fourier transform of \eqref{transfEcfieldgen} and integrating with $\xi(\omega)/\mathcal{E}(\omega)$ as before, we obtain
\begin{align}
    % a_{\xi}^\dagger &= \frac{1}{2\pi}\int_0^\infty d\omega\xi(\omega)\int d\nu \frac{\mathcal E(\nu)}{\mathcal E(\omega)} g(-\omega,\nu) c^{\regi^{\scriptstyle\dagger}}(\nu) \label{axiagen} \\
    % &+ \frac{1}{2\pi}\int_0^\infty d\omega\xi(\omega)\int d\nu \frac{\mathcal E(\nu)}{\mathcal E(\omega)} h(-\omega,\nu) d^{\regii^{\scriptstyle\dagger}}(\nu), \nonumber
    a_{\xi}^\dagger &= \frac{1}{2\pi}\int_0^\infty d\omega\int_{0}^{\infty} d\nu\xi(\omega) \frac{\mathcal E(\nu)}{\mathcal E(\omega)} 
        \notag \\
        &\ \ \ \,  \biggl(g(-\omega,\nu) a^{\regi^{\scriptstyle\dagger}}(\nu) + g(-\omega,-\nu) a^{\regi}(\nu)
         \notag \\
        &\ \   + h(-\omega,\nu) b^{\regii^{\scriptstyle\dagger}}(\nu) + h(-\omega,-\nu) b^{\regii}(\nu)
        \biggr), \label{axiagen}
\end{align}
where $g(\omega,\nu)$ and $h(\omega,\nu)$ are found from $g(x,y)$ and $h(x,y)$ by Fourier transforms w.r.t. $x$ and $y$.

The forward photon number is 
\begin{align}\label{forwphotn}
    \expval{n} &= \sum_\xi \bra{0^\regi}\bra{0^{\regii}} a_\xi^\dagger a_\xi \ket{0^\regi}\ket{0^{\regii}} \\
    &= \frac{1}{(2\pi)^2}\int_0^\infty d\nu \int_0^\infty d\omega \frac{\nu}{\omega} \lp[ |g(\omega,\nu)|^2 + |h(\omega,\nu)|^2 \rp], \nonumber
\end{align}
where we have used $\mathcal E^2(\omega)\propto |\omega|$. The total photon number, on the other hand, may be written as 
\begin{align}
    \langle n_{\text{tot}}\rangle = \langle n\rangle + \sum_{\xi} \bra{0^\regi}\bra{0^{\regii}} b_\xi^\dagger b_\xi \ket{0^\regi}\ket{0^{\regii}},
\end{align}
where the last term is to be interpreted as the expected number of backward-propagating photons in the initial vacuum state. 
The form of $\raisemath{-2pt}{\rule{0pt}{10pt}}\smash{b_{\xi}^{\dag}}$, which can be obtained from \eqref{transfEdfieldgen} in the same way as $\raisemath{-2pt}{\rule{0pt}{10pt}}\smash{a_{\xi}^{\dag}}$ from \eqref{transfEcfieldgen}, is identical to the right-hand side of \eqref{axiagen}, only with $g(-\omega,\nu)$ and $h(-\omega,\nu)$ interchanged. It then follows that the total photon number becomes 
\begin{align}\label{ntotn}
    \langle n_{\text{tot}}\rangle = 2\langle n\rangle.
\end{align} 
By substituting \eqref{gisak} and using properties of $f(x,y)$, one can then estimate $\expval{n}$, and at the same time, $\langle n_{\text{tot}}\rangle$, using \eqref{forwphotn}.

Before concretizing the gradual removal, we calculate a fidelity estimate for a general model \eqref{transffieldgen}, valid for $\expval{n} \ll 1$. Following the derivation for $F_{\xi}$ in the main text, the more general expression takes the form
\begin{align}
    F_{\xi}  = |\langle \xi, \theta_+T_1\xi\rangle| = |\langle T_1^{\dag}\xi,\xi\rangle|,
\end{align}
where $\theta_+(\omega) = \theta(+\omega)$. The last equality holds since $\xi(\omega)=0$ for $\omega <0 $. Comparing \eqref{axiagen} with \eqref{inverseBogo} and  \eqref{ourBogo}, we find that 
\begin{align}
    (T_1^{\dag}\xi)(\nu) = 
    \frac{1}{2\pi} \int_{0}^{\infty}d\omega \frac{\mathcal E(\nu)}{\mathcal E(\omega)}\xi(\omega) g(-\omega,\nu), \ \ \nu > 0.\label{T_dag_general_trans}
\end{align}
Using \eqref{gisak}, the fidelity then takes the form
\begin{align}
    F_{\xi} = |\langle T_1^{\dag}\xi, \xi\rangle| = |\langle \xi_{-}^{+} + \bar{\xi}, \xi\rangle| = |\langle \xi, \xi_{-}^{+} + \bar{\xi}\rangle|,  \label{F_xi_supplemental}
\end{align}
where 
\begin{align}
    \bar{\xi}(\nu) = \frac{1}{2\pi} \int_{0}^{\infty}d\omega \frac{\mathcal E(\nu)}{\mathcal E(\omega)} \xi(\omega)f(-\omega,\nu).
\end{align}
We have
\begin{align}
    \langle\xi,\bar{\xi}\rangle &= \frac{1}{2\pi}\int_{0}^{\infty}d\omega \int_{0}^{\infty}d\nu \frac{\mathcal E(\nu)}{\mathcal E(\omega)} \xi(\omega)f(-\omega,\nu) \xi^*(\nu) 
    \notag \\
    % &\leq \pi||\xi||^2\langle n\rangle,
    &= 2\pi \int dx\int dy \xi(x)f(x,y)\tilde{\xi}^*(y).
    \label{correction_term_F_xi}
\end{align}
As in the main article, we conclude that the fidelity is given by ``the fraction of the photon'' located to the left of the transition region for $t = 0$, but with a correction term that describes in detail how the part of the photon in the transition region is cut.

\subsection{Electromagnetic model}

Here, we consider a reflector which is gradually turned off over some transition region $-T\leq t<0$. We do so by considering an infinitely thin,  dielectric slab, with permittivity \cite{braunstein2005}
\begin{align}
    \epsilon(x,t) = 1 + \kappa(t)\delta(x). \label{eq:epsilon}
\end{align}
Here \(\kappa(t)\) is equal to a large value $\kappa_0$ for $t\leq -T$, and zero for $t\geq 0$. In the transition region $-T<t<0$, $\kappa(t)$ is assumed to be continuous everywhere, with continuous first derivative.

The electric field is assumed to point in the $\vekh y$-direction, while the magnetic field is in the $\vekh z$-direction. In each region $\regi$ ($x<0$) and $\regii$ ($x>0$) we decompose the fields in their forward- ($a$) and backward-propagating ($b$) parts:
\begin{subequations}    \label{E_and_B_field_appendix}
\begin{align}
    E(x,t) &= 
    \begin{cases}
        E^{\regi}_{a}(t-x) + E^{\regi}_{b}(t+x), & x<0, \\
        E^{\regii}_{a}(t-x) + E^{\regii}_{b}(t+x), & x>0,
    \end{cases} \\
    B(x,t) &=
    \begin{cases}
        E^{\regi}_{a}(t-x) - E^{\regi}_{b}(t+x), & x<0, \\
        E^{\regii}_{a}(t-x) - E^{\regii}_{b}(t+x), & x>0.
    \end{cases}  
\end{align}
\end{subequations}

From Maxwell's equations, we find the boundary conditions
\begin{subequations}\label{maxwbound}
\begin{align}
    E(0^+,t) &= \tilde{E}(t) = E(0^-,t),
    \label{eq:continuity} \\
    B(0^+,t) - B(0^-,t) &= -\kappa(t)\partial_t \tilde{E}(t) - \dot{\kappa}(t) \tilde{E}(t),   \label{eq:from_Maxwell}
\end{align} 
\end{subequations}
where $\tilde{E}(t)$ is the electric field inside the slab. In obtaining the boundary conditions we have introduced a high-frequency cutoff for the quantum field integral expressions. While the cutoff can be arbitrarily high, it has been fixed to make the zero slab thickness meaningful. Being bandlimited, it is clear that the quantum fields are continuous in space and time.

It is natural to view $E_a^\regi(t-x)$ and $E_b^\regii(t+x)$ as the independent input fields, as they propagate from $\pm\infty$ towards the reflector. For ease of notation we denote them $E^\regi(t-x)\equiv E_a^\regi(t-x)$ and $E^\regii(t+x)\equiv E_b^\regii(t+x)$, just as we have done in the main text. Eq. \eqref{eq:continuity} can be used to express the output fields from the input fields:
\begin{subequations}    \label{eq:reflected_fields_removed}
    \begin{align}
        E^{\regi}_{b}(t) &= \tilde{E}(t) - E^{\regi}(t), \\
        E^{\regii}_{a}(t) &= \tilde{E}(t) - E^{\regii}(t).
    \end{align}
\end{subequations}
By eliminating the output fields, \eqref{eq:from_Maxwell} becomes
\begin{align}
    % \dot{\tilde{E}}(t) + \left[\frac{\dot{\kappa}(t)}{\kappa(t)} + \frac{2}{\kappa(t)}\right]\tilde{E}(t) = \frac{2}{\kappa(t)} \left[E^{\regi}(t) + E^{\regii}(t)\right],  
    \kappa(t)\dot{\tilde{E}}(t) + \left[\dot{\kappa}(t) + 2\right]\tilde{E}(t) = 2\left[E^{\regi}(t) + E^{\regii}(t)\right],  
    \label{eq:Etilde_before_G}
\end{align}
which is the equation of motion for the electric field in the slab, given an input field excitation.

Dividing \eqref{eq:Etilde_before_G} by $\kappa(t)$ and multiplying with 
\begin{align}
    G(t) &= \exp[\int_{-T}^{t}d\tau\lp(\frac{\dot{\kappa}}{\kappa} + \frac{2}{\kappa} \rp)] 
    \notag \\
    &= 
    \frac{\kappa(t)}{\kappa_{0}} \exp\lp[2\int_{-T}^{t}\frac{d\tau}{\kappa(\tau)}\rp] 
\end{align}
gives 
\begin{align}
    \frac{d}{dt}\left(G(t)\tilde{E}(t)\right) = \frac{2G(t)}{\kappa(t)}\left[E^{\regi}(t) + E^{\regii}(t)\right].   \label{eq:Etilde_before_integration}
\end{align}
Integrating from $-\infty$ to $t$, we obtain the solution
\begin{align}
    \tilde{E}(t) &= \frac{2}{\kappa(t)}e^{-2K(t)} \int_{-\infty}^{t}dt' e^{2K(t')} \left[ E^{\regi}(t') + E^{\regii}(t') \right]  \notag \\
    &= \frac{2}{\kappa(t)} \int_{-\infty}^{t}dt' \exp\lp[-2\int_{t'}^t \frac{d\tau}{\kappa(\tau)} \rp] \left[ E^{\regi}(t') + E^{\regii}(t') \right] \label{eq:tildeE_bef}
\end{align}
where we have defined 
\begin{align}
    K(t) \equiv \int_{-T}^{t}\frac{d\tau}{\kappa(\tau)}.
\end{align}

The solution \eqref{eq:tildeE_bef} shows that the field in the slab at time $t$ depends on the input fields at earlier times $t'\leq t$, in accordance with causality. It might appear surprising that there is a memory in the system; one would perhaps expect that the field in the slab at $t$ is only dependent on the input field at $t$. However, even though the thickness of the slab tends to zero, the permittivity is correspondingly large, which leads to large Fresnel reflection coefficients at the interfaces. The presence of multiple reflections therefore gives stored energy.

For $t<-T$ the mirror is constantly on, characterized by $\kappa(t) = \kappa_0$. For a situation where $\kappa(t) = \kappa_0$ all the time, we can evaluate \eqref{eq:tildeE_bef} and calculate the Fourier transform with the help of the convolution theorem, to obtain
\begin{equation}\label{Etildom}
    \tilde E(\omega) = \frac{E^\regi(\omega) + E^\regii(\omega)}{1-i\omega\kappa_0/2}.
\end{equation}
Eq. \eqref{Etildom} is valid both for quantum fields and classical fields. Considering an incident, classical wave from the left, and using \eqref{Etildom} and \eqref{eq:reflected_fields_removed}, we obtain the transmission coefficient
\begin{align}\label{transcoeff}
    \mathcal T \equiv \frac{E_a^\regii(\omega)}{E^{\regi}(\omega)} = \frac{1}{1-i\omega\kappa_0/2}. 
\end{align}

The mirror is perfectly removed at \(t=0\). 
Since the fields \eqref{E_and_B_field_appendix} can be expressed solely in terms of the continuous input fields and $\tilde{E}(t)$, there will be temporal continuity across $t=0$, provided $\tilde{E}(t)$ is continuous there as well. Evaluating \eqref{eq:Etilde_before_G} at \(t=0^{\pm}\) leads to
\begin{align}
    \tilde{E}(0^{\pm}) = \frac{2}{2+\dot{\kappa}(0^{\pm})} \lp[E^{\regi}(0^{\pm}) + E^{\regii}(0^{\pm})\rp].
\end{align}
Hence, ensuring temporal continuity at $t=0$ amounts to ensuring continuity of $\dot{\kappa}(t)$ there.  
Since the mirror is absent for $t>0$, we have $\dot{\kappa}(0^+)=0$ by construction, and so we must thus require $\dot{\kappa}(0^-)=0$. This does mean that \(K(0)\) is divergent; however, this is no issue for the well-definedness of the expression for \(\tilde{E}(t)\) in \eqref{eq:tildeE_bef}. We thus get temporal continuity at $t=0$ by an appropriate choice of $\kappa(t)$. This is valid for any finite (but possibly, arbitrarily small) mirror removal time $T$.

For $t>0$ we label the forward-propagating field by $E_a(t-x)$ and the backward-propagating field by $E_b(t+x)$, just as in the main article. The temporal continuity at $t=0$ gives the connection
\begin{subequations}    \label{eq:general_new_fields_reflected_backward}
\begin{align}
    E_a(x) &= \theta(x) E_{a}^{\regi}(x) + \theta(-x)E_{a}^{\regii}(x), \\
    E_b(x) &= \theta(-x) E_{b}^{\regi}(x) + \theta(x)E_{b}^{\regii}(x),
\end{align}
\end{subequations}
which after inserting \eqref{eq:reflected_fields_removed} gives the transformation
\begin{subequations}    \label{eq:transformation_final}
    \begin{align}
        E_a(x) &= \theta(x) E^{\regi}(x) + \theta(-x)\big[\tilde{E}(x) - E^{\regii}(x)\big], \label{eq:forward_transformation_final}\\
        E_b(x) &= \theta(-x) \big[\tilde{E}(x) - E^{\regi}(x)\big] + \theta(x)E^{\regii}(x).
    \end{align}
\end{subequations}

Although not needed here, it is also possible to formulate the inverse transformation, i.e., given $E_a(x)$ and $E_b(x)$, determine $E^\regi(x)$ and $E^\regii(x)$. Start with \eqref{eq:transformation_final}, which can be written as 
\begin{subequations}    \label{eq:transfinv1}
    \begin{align}
        E^\regi(x) &= \theta(x) E_a(x) + \theta(-x)\big[\tilde{E}(x) - E_b(x)\big], \\
        E^\regii(x) &= \theta(-x) \big[\tilde{E}(x) - E_a(x)\big] + \theta(x)E_b(x).
    \end{align}
\end{subequations}
Consider \eqref{eq:Etilde_before_G} for $t<0$. According to \eqref{eq:transfinv1} we have $E^\regi(x) + E^\regii(x) = 2\tilde E(x)-E_a(x)-E_b(x)$, for $x<0$. Substituting into 
\eqref{eq:Etilde_before_G} we find
\begin{align}
    \kappa(t)\dot{\tilde{E}}(t) + \left[\dot{\kappa}(t) - 2\right]\tilde{E}(t) = -2 \left[E_a(t) + E_b(t)\right]. \label{invtranstildeE}
\end{align}
This is the same equation of motion as for the direct case, except that the sign of $\kappa(t)$ has been flipped. In other words, the inverse transformation $E_a, E_b$ $\mapsto$ $E^\regi, E^\regii$ has the same form as the direct transformation, the only difference being the sign of $\kappa(t)$.
 
\subsection{Commutator relations}
In the following, for ease of notation, we have normalized the field such that $\mathcal{E}(\omega)=\sqrt{\omega}$. In order for \eqref{eq:transformation_final} with \eqref{eq:tildeE_bef} to describe a unitary transformation, we should verify that 
\begin{subequations}
    \begin{align}
        & \left[E_a(x), E_b(y)\right] = 0, \label{eq:E_u_E_v_verify_me}\\
        & \left[E_a(x), E_a(y)\right] = \left[E_b(x), E_b(y)\right] = 2\pi i\delta'(x-y),   \label{eq:E_u_E_u_verify_me}
    \end{align}
\end{subequations}
using the corresponding commutators for \(E^{\regi}\) and \(E^{\regii}\). Since our transformation is induced by a physical device governed by Maxwell's equations, we expect that this is the case.

\begin{widetext}
To verify the commutator relations, we begin with
\begin{align}
    \left[E_a(x), E_b(y)\right] &= \theta(-x)\theta(-y)\left( \bigl[\tilde{E}(x), \tilde{E}(y)\bigr] - \bigl[\tilde{E}(x), E^{\regi}(y)\bigr] - \bigl[ E^{\regii}(x), \tilde{E}(y) \bigr]\right).   \label{eq:Ec_Ed_should_vanish}
\end{align}
From \eqref{eq:tildeE_bef} we have 
\begin{align}
    \bigl[\tilde{E}(x), \tilde{E}(y)\bigr] &= \frac{4}{\kappa(x)\kappa(y)}e^{-2K(x)-2K(y)} \int_{-\infty}^{x}dx'\int_{-\infty}^{y}dy' e^{2K(x')+2K(y')} 4\pi i\delta'(x'-y')
    \notag \\
    &= \frac{16\pi i}{\kappa(x)\kappa(y)}e^{-2K(x)-2K(y)} \int dx'\int dy' \theta(x-x')\theta(y-y') e^{2K(x')+2K(y')} \delta'(x'-y')
    \notag \\
    &= \frac{16\pi i}{\kappa(x)\kappa(y)}e^{-2K(x)-2K(y)} \int dx'\int dy' e^{2K(x')} \theta(x-x')\frac{d}{dy'}\lp[\theta(y-y') e^{2K(y')}\rp] \delta(x'-y'),    \label{eq:first_E_tildE_aomm_calc}
\end{align}
which by using that $d\theta(x)/dx = \delta(x)$ becomes \footnote{After evaluating the integrals in \eqref{eq:first_E_tildE_aomm_calc} one is left with a term proportional to $1/2-\theta(0)$. Demanding that \eqref{eq:first_E_tildE_aomm_calc} is antisymmetric under the interchange $x\leftrightarrow y$ forces ${\theta(0) = 1/}2$, resulting in \eqref{eq:final_E_tildE_aomm_calc}.}
\begin{align}
    \bigl[\tilde{E}(x), \tilde{E}(y)\bigr] &= -\frac{16\pi i}{\kappa(x)\kappa(y)}e^{-2K(x)-2K(y)} \lp[\theta(x-y)e^{4K(y)} - \frac{1}{2}e^{4K(x')}\big\vert_{\min\{x,y\}}\rp]
    \notag \\
    &= \frac{8\pi i}{\kappa(x)\kappa(y)}\lp[\theta(y-x)e^{2K(x)-2K(y)} - \theta(x-y)e^{2K(y)-2K(x)}\rp].  \label{eq:final_E_tildE_aomm_calc}
\end{align}
Similarly we find
\begin{subequations}\label{eq:commtIII}
\begin{align}
    \bigl[\tilde{E}(x), E^{\regi}(y)\bigr] 
    &= \frac{4\pi i}{\kappa(x)}e^{-2K(x)}\int_{-\infty}^{x}dx' e^{2K(x')}\delta'(x'-y) = \frac{4\pi i}{\kappa(x)}\delta(x-y) -\frac{8\pi i}{\kappa(x)\kappa(y)} \theta(x-y)e^{2K(y)-2K(x)}, \\ 
    \bigl[E^{\regii}(x), \tilde{E}(y) \bigr] 
    &= -\frac{4\pi i}{\kappa(y)}\delta(x-y) +\frac{8\pi i}{\kappa(x)\kappa(y)} \theta(y-x)e^{2K(x)-2K(y)}.
\end{align}
\end{subequations}
Substituting \eqref{eq:final_E_tildE_aomm_calc} and \eqref{eq:commtIII} into \eqref{eq:Ec_Ed_should_vanish} shows that \([E_a(x), E_b(y)]=0\).

Lastly, we want to show that \eqref{eq:E_u_E_u_verify_me} is satisfied. This follows from the fact that \eqref{eq:E_u_E_v_verify_me} holds, since
\begin{align}
    \left[E_a(x), E_a(y)\right] &= 
    2\pi i\delta'(x-y) + \theta(-x)\theta(-y)\bigl(\left[\tilde{E}(x), \tilde{E}(y)\right] - \left[\tilde{E}(x), E^{\regii}(y)\right] - \left[E^{\regii}(x), \tilde{E}(y)\right] \bigr)
    \notag \\
    &= 2\pi i\delta'(x-y) + \left[E_a(x), E_b(y)\right] 
    = 2\pi i\delta'(x-y), \label{eq:E_a_E_b_first_calc}
\end{align}
where we have used \eqref{eq:Ec_Ed_should_vanish} and the fact that $[\tilde{E}(x), E^{\regii}(y)] = [\tilde{E}(x), E^{\regi}(y)]$, which follows from \eqref{eq:tildeE_bef}.
\end{widetext}

\subsection{Photon number estimation}
To estimate the photon number in the initial vacuum state, we use \eqref{forwphotn} with \eqref{gisak}. From \eqref{transffieldgen}, \eqref{gisak}, \eqref{eq:tildeE_bef}, and \eqref{eq:transformation_final} we have for $-\infty<x<0$ and $-\infty<y<x$ that 
\begin{align}\label{fident}
    f(x,y) = \frac{2}{\kappa(x)} \Gamma(x,y),
\end{align}
with 
\begin{align}
    \Gamma(x,y) = \exp\lp(-2\int_y^x \frac{d\tau}{\kappa(\tau)} \rp).
\end{align}
In the following we will utilize the fact that terms of the form \(\delta(x-y)\) in $g(x,y)$ or $h(x,y)$ will become \(\delta(\nu+\omega)\) in the Fourier domain, and these will not contribute in \eqref{forwphotn}. A finite photon number is obtained if these functions otherwise behave sufficiently nicely, such that the Fourier domain integral \eqref{forwphotn} is finite. The function $f(x,y)$ is however not particularly nice, as $\theta(-x)f(x,y)$ tends to $\theta(-x)\delta(y)$ as $x\to 0$. Nevertheless, it turns out that this behavior actually is quite useful for us, since this delta function combines with the term $\theta(x)\delta(x-y)$ in \eqref{gIisak}, effectively extending the support of the $\theta(x)$ function to small negative values of $x$. In \eqref{gIIisak} we similarly get a partial cancellation. Therefore, we expect that the combination in \eqref{gisak} provides the required smoothening of the instantaneous mirror removal in the main article, giving a bounded photon number.

We begin by extending the support of \({\theta(x)\delta(x-y)}\) in \eqref{gIisak} by ``extracting'' \(\theta(-x)\delta(x-y)\) from \(f(x,y)\). 
Noting that
\begin{equation}
    f(x,y) = -\frac{d\Gamma(x,y)}{dx},    
\end{equation}
integration by parts leads to
\begin{align}
    f(\omega,\nu) &= \int_{-\infty}^0 dx e^{i(\omega+\nu)x} \notag \\
    &+ i\omega\int_{-\infty}^{0} dx \int_{-\infty}^{x} dy \Gamma(x,y) e^{i\omega x+i\nu y}.   \label{eq:f_nu_omega_partial_first}
\end{align} 
Here, the boundary term at \(x=0\) vanishes since $\kappa(\tau)$ and $\dot\kappa(\tau)$ tend to zero as $\tau\to 0$. The first term in \eqref{eq:f_nu_omega_partial_first} will combine with the term \(\theta(x)\delta(x-y)\) in \eqref{gIisak}, creating a term \(\delta(x-y)\) that does not contribute to \eqref{forwphotn}. In \eqref{gIIisak}, we will have an exact cancellation. In both cases, \(g(\omega,\nu)\) and \(h(\omega, \nu)\) effectively take the form as the last term in \eqref{eq:f_nu_omega_partial_first}.

Considering this last term further, we now split the \(x\)-integral into $I_1 + I_2$, where
\begin{align}
    I_1 &= i\omega\int_{-\infty}^{-T} dx \int_{-\infty}^{x} dy e^{-\frac{2}{\kappa_0}(x-y)} e^{i\omega x + i\nu y} \nonumber\\ 
    &=
    \frac{i\omega e^{-i(\nu + \omega)T}}{\frac{2}{\kappa_0} + i \nu} \left( \pi \delta(\omega+\nu) - i \frac{1}{\omega+\nu}\right),
\end{align}\label{I1res}
and
\begin{align}
    I_2 &= i\omega\int_{-T}^{0} dx \int_{-\infty}^{x} dy \Gamma(x,y) e^{i\omega x + i\nu y} 
    % \nonumber
    \label{eq:I2_alternative_form} \\
    &= i\omega \int_{-T}^{0}dx\int_{0}^{\infty}du \exp[-2\int_{x-u}^{x}\frac{dt}{\kappa(t)}]e^{i(\nu+\omega)x - i\nu u}.
    \nonumber
\end{align}
Integration by parts in $x$ and substituting back to \({y=x-u}\) gives 
\begin{align}
    I_1 + I_2 = \frac{\omega}{\omega+\nu}\tilde{\Gamma}(\omega,\nu) + \frac{i\pi\omega e^{-i(\omega+\nu)T}}{\frac{2}{\kappa_0} + i \nu}  \delta(\omega+\nu), \label{I1pI2}
\end{align}
where
\begin{align}
    \tilde{\Gamma}(\omega,\nu) = \int_{-T}^{0}dx \int_{-\infty}^{x}dy \tilde{\Gamma}(x,y)e^{ix\omega +iy\nu},
\end{align}
and 
\begin{align}
   \tilde{\Gamma}(x,y) = \left(\frac{2}{\kappa(x)} - \frac{2}{\kappa(y)}\right)\Gamma(x,y). \label{eq:tilde_Gamma_xy}
\end{align}
Since terms proportional to $\delta(\omega+\nu)$ do not contribute to \eqref{forwphotn}, we end up with 
\begin{align}
    \langle n\rangle = \frac{2}{(2\pi)^2}\int_{0}^{\infty}d\nu \int_{0}^{\infty}d\omega \frac{\nu \omega}{(\nu + \omega)^2} |\tilde{\Gamma}(\omega,\nu)|^2.
\end{align}
This is our final exact form for the photon number, and we now proceed to give a bound for this expression. 

Note first that \(\nu\omega/(\omega+\nu)^2\leq 1/4\), which means that
\begin{align}
    \langle n\rangle &\leq \frac{1}{2(2\pi)^2}\int_{-\infty}^{\infty}d\nu \int_{-\infty}^{\infty}d\omega |\tilde{\Gamma}(\nu,\omega)|^2
    \notag \\
    &= \frac{1}{2} \int_{-T}^{0}dx \int_{-\infty}^{x}dy |\tilde{\Gamma}(x,y)|^2 ,   \label{eq:bound_n}
\end{align}
where we have extended the domain of the \(\nu\)- and \({\omega\text{-integral}}\) and then used the Plancherel theorem.

We now consider an explicit choice of \(\kappa(t)\), namely 
\begin{align}
    \kappa(t) = \begin{cases}
        \kappa_0\frac{t^2}{T^2}, &-T\leq t\leq 0, \\
        \kappa_0, &t < -T.   
    \end{cases} \label{eq:kappa_explicit}
\end{align}
This choice of \(\kappa(t)\) means that \(\Gamma(x,y)\) takes the form 
\begin{align}
    \Gamma(x,y) = \begin{cases}
        \exp(\frac{2}{\kappa_0}\left(\frac{T^2}{x}+y+2T\right)), &y\leq -T, \\
        \exp(\frac{2}{\kappa_0}\left(\frac{T^2}{x} - \frac{T^2}{y}\right)), &y>-T.   \label{eq:explicit_Gamma}
    \end{cases}
\end{align}
We then split up the \(y\)-integral in \eqref{eq:bound_n} into two regions, \(y\leq -T\) and \(y>-T\). Note that the \(y\)-variable corresponds to the integration variable in the expression for \(\tilde{E}(t)\) (see \eqref{eq:tildeE_bef}). Hence, these two regions can be thought of as the contribution from the distant past and the transition region, respectively. After some algebra, we end up with 
\begin{align}\label{nbound}
    \langle n\rangle \leq \frac{\kappa_0}{4T} + \frac{\kappa_0^2}{16T^2},
\end{align}
where the first term stems from the transition region, while the last from the distant past. With the result \eqref{nbound}, we also have a bound for the total photon number $\langle n_{\text{tot}}\rangle$ since $\langle n_{\text{tot}}\rangle = 2 \langle n\rangle$.

To interpret this result for our case, we must first pick a sufficiently large, initial reflector strength $\kappa_0$, such that the transmissivity for the incident photon is very low. Let $\omega_0$ denote the central frequency of the incident photon. From \eqref{transcoeff} we must have $\omega_0\kappa_0 \gg 1$ to obtain a small $|\mathcal T|^2$. 
Together with \eqref{nbound} we then conclude that the requirement for both a low transmissivity and a small photon number is
\begin{align}
\frac{1}{\omega_0} \ll \kappa_0 \ll T.
\end{align}
As a numerical example, to achieve a transmissivity $|\mathcal T|^2 = 10^{-4}$, we need $\omega_0\kappa_0=200$. Then the photon number is small provided $T \gg 200/\omega_0$. For a photon with frequency $\omega_0/2\pi=10^{15}$~Hz, the transition region of the reflector can be as small as $T \sim 10^{-14}$~s before the photon number becomes comparable to 1. 

% \bibliography{single_photon, appendix_references}
% \bibliography{references}
% \input{manual_references.bbl}  %% Copy of references.bib but with the numbering from before referencing supplemental (which should only appear in the main text alone) skipped one ahead.

%apsrev4-2.bst 2019-01-14 (MD) hand-edited version of apsrev4-1.bst
%Control: key (0)
%Control: author (72) initials jnrlst
%Control: editor formatted (1) identically to author
%Control: production of article title (-1) disabled
%Control: page (0) single
%Control: year (1) truncated
%Control: production of eprint (0) enabled
%

%%%%% Manual numbering:

\end{document}